\documentclass[a4paper,11pt]{article}
\usepackage{graphicx,amssymb,amsmath,bm,latexsym,color,epsf,cite,comment,breqn}
\makeatletter
\@addtoreset{equation}{section}

\makeatother
\newcommand{\bea}{\begin{eqnarray}}
\newcommand{\eea}{\end{eqnarray}}
\newcommand{\vs}[1]{\vspace{#1 mm}}
\newcommand{\hs}[1]{\hspace{#1 mm}}
\renewcommand{\a}{\alpha}
\renewcommand{\b}{\beta}

\newcommand{\e}{\epsilon}

\renewcommand{\t}{\theta}

\newcommand{\la}{\lambda}
\newcommand{\pa}{\partial}
\newcommand{\nn}{\nonumber\\}

\begin{document}

%\begin{flushright}
% \\
%\today
%\end{flushright}

\begin{center}
{\Large\bf An Observation on the Beta Functions in Quadratic Gravity}
\vs{10}

{\large
Hikaru Kawai$^{a,b,}$\footnote{e-mail address: hikarukawai@phys.ntu.edu.tw}
and
Nobuyoshi Ohta$^{c,d,}$\footnote{e-mail address: ohtan.gm@gmail.com}
} \\
\vs{5}

$^a${\em Department of Physics and Center for Theoretical Physics, National Taiwan University, Taipei 106, Taiwan}

$^b${\em Physics Division, National Center for Theoretical Sciences, Taipei 106, Taiwan}
\vs{2}

$^c${\em Research Institute for Science and Technology, Kindai University, Higashi-Osaka, \\
Osaka 577-8502, Japan}

$^d${\em School of Physics, Korea Institute for Advanced Study, 85 Hoegi-ro Dongdaemun-gu, \\
Seoul 02455, Korea}

\vs{10}
%%%%%%%%%%%%%%%%%%%%%%%%%%%%%%%%
{\bf Abstract}
\end{center}

We study the beta functions for the dimensionless couplings in quadratic curvature gravity,
and find that there is a simple argument to restrict the possible form of the beta functions as
derived from the counterterms at an arbitrary loop.
The relation to the recent different results on beta functions is also commented on.

\vs{10}
\setcounter{footnote}{0}

%%%%%%%%%%%%%%%%%%%%%%%%%%%%
\section{Introduction}
%%%%%%%%%%%%%%%%%%%%%%%%%%%%

The perturbative calculation of beta functions in quadratic gravity has a long history.
The action of our concern is
\bea
S=\int d^4 x \sqrt{-g} \left[\frac{1}{2\la} C_{\mu\nu\rho\la}^2 +\frac{1}{\xi} R^2
-\frac{1}{\kappa} E +\frac{M_P^2}{2} R +\rho \right],
\label{action}
\eea
where $M_P$ is the Planck mass, and $C_{\mu\nu\rho\la}$ is the Weyl tensor with
$C_{\mu\nu\rho\la}^2= R_{\mu\nu\rho\la}^2-2 R_{\mu\nu}^2 +\frac13 R^2$  and
$E=R_{\mu\nu\rho\la}^2-4 R_{\mu\nu}^2 + R^2$.

Traditionally the beta functions for the dimensionless couplings are defined by
\bea
\b_\la^o = \mu \frac{d\la}{d\mu}, \qquad
\b_\xi^o = \mu \frac{d\xi}{d\mu}, \qquad
\b_\kappa^o = \mu \frac{d\kappa}{d\mu},
\eea
where $\la, \xi$ and $\kappa$ are the renormalized couplings and $\mu$ is the renormalization scale.
In particular, in the minimal subtraction scheme, $\mu$ is the momentum scale used in the dimensional regularization
\bea
S=\int d^d x \sqrt{-g} \mu^{-2\epsilon}\left[\frac{1}{2\la} C_{\mu\nu\rho\la}^2 +\frac{1}{\xi} R^2
-\frac{1}{\kappa} E +\frac{M_P^2}{2} R +\rho \right]+\text{counterterms}.
\label{Daction}
\eea
Here, $\epsilon=\frac{4-d}{2}$ and the counterterms correspond to the minimal subtraction.
Note that the beta functions depend on the renormalization schemes. In general, 
however, they are universal up to two-loop order for mass-independent renormalization schemes.

The first attempt to derive the beta functions for $\la$ and $\xi$ was made in \cite{JT},
but this does not have contributions from additional ghosts.
The result was corrected in \cite{FT}, with some further corrections in \cite{AB}.
The final correct answer in this definition of the beta functions is
\bea
\b_\la^o &=& -\frac{1}{(4\pi)^2} \frac{133}{10}\la^2, \nn
\b_\xi^o &=& -\frac{1}{(4\pi)^2} \frac{5(72\la^2 -36 \la\xi +\xi^2)}{36} .
\label{old}
\eea
Since then, the result is confirmed by several works~\cite{CP,Niedermaier,SGRZ,BS2,OP}.
It was noted that the beta function for the coefficient of the Gauss-Bonnet term $\kappa$ does not depend on
$\kappa$ itself because it is a topological term which does not contribute to quantum effects~\cite{FOP}.
Thus it is proportional to $\kappa^2$ at the fixed point of other couplings, and it is either asymptotically
free or not depending on the sign of the proportionality constant.
It is claimed that the other couplings are asymptotically free, but there is restriction on these
couplings $\xi>0.5487 \lambda>0$ to be asymptotically free~\cite{KO}.

This result means that the simple pole part of the counterterms are given as
\bea
\frac{\mu^{-2\e}}{4\e } \frac{\b_\la^{o}}{\la^2} C^2 +\frac{\mu^{-2\e}}{2\e} \frac{\b_\xi^{o}}{\xi^2} R^2,
\label{beta}
\eea
in the dimensional regularization, $\frac{\mu^{-2\e}}{\e}$
corresponding to $\log\left(\frac{\Lambda_{UV}^2}{\mu^2}\right)$ in the cutoff regularization.
See, for example, the review in \cite{Ohta}.

In a recent paper~\cite{BDMP}, it has been claimed that the above results on the beta functions
of the dimensionless couplings $\la$ and $\xi$ do not describe the running of the couplings.
The beta functions obtained in Ref.~\cite{BDMP} are given as
\bea
\b_\la^n &=& -\frac{1}{(4\pi)^2} \frac{(1617\la -20 \xi)\la}{90}, \nn
\b_\xi^n &=& -\frac{1}{(4\pi)^2} \frac{ \xi^2 -36 \la\xi -2520 \la^2}{36},
\label{new}
\eea
where $\la$ and $\xi$ are the couplings read from the two-point functions with typical momentum $p$,
and the above beta functions are defined by the $p$-derivatives of these couplings.
The $p$-dependence is usually the same as the $\mu$-dependence, because for dimensional reasons they occur
as $\log(p/\mu)$ in the theory with a single momentum scale $p$. However, they claim
that in the limit of $p\gg M_P$, the propagator leads to infrared (IR) divergence which must be regularized,
and this gives additional $\log(p^2/m^2)$ dependence with IR regulator mass $m$.
(Similar infrared divergence was first noted in \cite{SSV}.)
They also claim that it is these beta functions that describe the ``physical'' running of
the couplings that are involved in the scattering amplitudes.
%Note that the definition of the beta functions are different from the above conventional ones.

Motivated by this result, we have studied the structure of the beta functions in perturbation theory
in the mass-independent renormalization scheme~\cite{Weinberg}, and point out that there is a simple argument
to restrict the possible form of the beta functions as derived from the counterterms at an arbitrary loop.
By considering the $\xi \to \infty$ limit, we argue that the conventional beta function for $\la$
should not depend on $\xi$.

In Ref.~\cite{SS}, similar $\xi\to\infty$ limit is considered in the theory without Einstein term.
The authors argue that the $\xi$ runs to infinity in the high energy limit according to the renormalization group
flow without Landau singularity by choosing suitable initial conditions, and then the conformal gravity can be defined
because the conformal mode decouples in the ultraviolet limit.
In contrast, we do not consider the running of the $\xi$ but just consider the $\xi\to\infty$ limit and show
that the scalar mode in the theory decouples in this limit and use this fact to restrict the general form
of the beta functions.
Also our result is not just restricted to one-loop; we argue that similar result is true also in higher loops.

Finally we also discuss the relation between various beta functions.

\section{Perturbation theory as a double expansion in $\lambda$ and $\xi$}

Applying the perturbation expansion to the action (\ref{action}), we see that
the $n$-loop correction is given by an $n$-th order homogeneous function of $\lambda$ and $\xi$.
Furthermore, as we will show, it is a degree $n$ homogeneous polynomial of $\lambda$ and $\xi$.
In other words, the perturbation series is nothing but a double expansion in $\lambda$ and $\xi$.

In order to see this, we rewrite (\ref{action}) using an auxiliary field $\varphi$ as
follows:\footnote{The Gauss-Bonnet term is ignored hereafter.}
\bea
S'=\int d^4 x \sqrt{-g} \left[ \frac{1}{2\la} C_{\mu\nu\rho\la}^2+ \left(\varphi +\frac{M_P^2}{2}\right) R
-\xi\varphi^2 +\rho \right].
\label{action2}
\eea
Then, by changing variables as
\bea
g'_{\mu\nu} = e^\phi g_{\mu\nu} , \qquad
e^\phi = \varphi + \frac{1}{2}M_P^2,
\eea
we have
\bea
S'=\int d^4 x \sqrt{-g'} \left[ \frac{1}{2\la} {C'}_{\mu\nu\rho\la}^2+ R'
- \frac{3}{2}{g'}^{\mu\nu} \partial_{\mu} \phi \  \partial_{\nu} \phi
-\xi\left(1-\frac{1}{2}M_P^2\  e^{-\phi}\right)^2 +\rho \ e^{-2\phi} \right],
\label{action3}
\eea
where ${C'}_{\mu\nu\rho\la}$ and $R'$ are the Weyl tensor and scalar curvature for ${g'}_{\mu\nu}$,
respectively.
Finally, by shifting $\phi$ as
\bea
\phi=\phi'+\log\frac{M_P^2}{2},
\eea
we have
\bea
S'=\int d^4 x \sqrt{-g'} \left[ \frac{1}{2\la} {C'}_{\mu\nu\rho\la}^2+ R'
- \frac{3}{2}{g'}^{\mu\nu} \partial_{\mu} \phi' \  \partial_{\nu} \phi'
-\xi(1-e^{-\phi'})^2 +\frac{4\rho}{M_P^4} \ e^{-2\phi'} \right].
\label{action4}
\eea

The important fact is that the Weyl gravity action, the first term in (\ref{action4}),
does not depend on the conformal factor of the metric. Therefore, if we decompose
the metric ${g'}_{\mu\nu}$ as
\bea
{g'}_{\mu\nu}=\eta_{\mu\nu}+\sqrt{\la} \bar{h}_{\mu\nu}+\frac{1}{4}\eta_{\mu\nu}h, \qquad
\bar{h}_{\mu\nu}\eta^{\mu\nu}=0,
\label{decompose}
\eea
the coefficients of the kinetic terms for $\bar{h}_{\mu\nu}$, $h$ and $\phi'$ are
numerical constants
and the other terms are associated with positive powers in $\sqrt{\la}$ and $\xi$.
This shows that the perturbation series is given by a power series of $\la$ and $\xi$.\footnote{
We note here that the cosmological constant term $\frac{4\rho}{M_P^4} \ e^{-2\phi'}$
in (\ref{action4}) is
a sort of mass term for $\phi'$ and it does not affect the beta functions in the minimal
subtraction scheme.}

\section{Smoothness of $\xi \to \infty$ limit and simple structure of $\beta_{\la}$ \label{xitoinf}}

Next we consider the limit where $\xi$ is taken to infinity.
First if we formally take this limit in (\ref{action}) (with the Gauss-Bonnnet term ignored),
we have
\bea
S_{\xi\to\infty}=\int d^4 x \sqrt{-g} \left[\frac{1}{2\la} C_{\mu\nu\rho\la}^2
+\frac{M_P^2}{2} R +\rho \right].
\label{Sinf}
\eea

From a quantum mechanical point of view, such a limit is not necessarily smooth.
In this case, however, we can show that this limit is indeed smooth as follows.
To see this, we consider the action (\ref{action4}), in which $\xi$ appears only in the fourth term.
It is clear that the field $\phi'$ is frozen to $\phi'=0$ in the $\xi \to \infty$ limit.
Then (\ref{action4}) is reduced to
\bea
S'_{\xi\to\infty}=\int d^4 x \sqrt{-g'} \left[ \frac{1}{2\la} {C'}_{\mu\nu\rho\la}^2+ R'
+\frac{4\rho}{M_P^4} \right],
\label{S'inf}
\eea
which is nothing but (\ref{Sinf}) with
\bea
g_{\mu\nu}=\frac{M_P^2}{2}g'_{\mu\nu}.
\eea

This argument can be made more rigorous by changing the variable in (\ref{action4}) as
\bea
\phi'=-\log(1-\psi) .
\eea
Then (\ref{action4}) becomes
\bea
S'=\int d^4 x \sqrt{-g'} \left[ \frac{1}{2\la} {C'}_{\mu\nu\rho\la}^2+ R'
- \frac{3}{2} \frac{1}{(1-\psi)^2} {g'}^{\mu\nu} \partial_{\mu} \psi \  \partial_{\nu} \psi
-\xi \psi^2 +\frac{4\rho}{M_P^4} (1-\psi)^2 \right].
\label{action5}
\eea
This shows that the mass of $\psi$ becomes infinite in the $\xi \to \infty$ limit
and that $\psi$ does not appear as an internal line. Therefore, $\psi$ decouples from the physical process.

This should be contrasted with the $\la \to \infty$ limit. In this limit, it is clear from (\ref{action4})
and (\ref{decompose}) that the traceless
mode of the metric, $\bar{h}_{\mu\nu}$, has large quantum fluctuations and couples with the
other degrees of freedom, $\phi'$ and $h$. Therefore, the system is a strongly coupled system.

In general, in the mass-independent renormalization scheme \cite{Weinberg}, the beta functions of dimensionless
parameters are independent of parameters with positive mass dimension.
Therefore, the beta functions for $\la$ and $\xi$ are independent of $M_P$ and $\rho$ and are functions
of $\la$ and $\xi$ only.

However, as we have seen, in the $\xi \to \infty$ limit, $\phi'$ decouples from the system and the
system is reduced to one with three parameters $\la$, $M_P$ and $\rho$. This means that the
beta function for $\la$ in a mass-independent scheme must be finite in the $\xi \to \infty$
limit. In other words, the
perturbation series for $\beta_{\la}$ must not have a term containing $\xi$.
At the 1-loop level, this is indeed the case, as shown in (\ref{old}).
The beta function of $\la$  has not been calculated beyond 1-loop, and it is
an interesting question to check whether this holds at higher loop levels.

Finally, we note that this can be also understood from the meaning of $\xi$ in (\ref{action4}).
Actually, it is clear that $\xi$ plays the roll of mass for $\phi'$. Therefore,
in the minimal subtraction scheme at least in the 1-loop level, it does not appear in the beta function of the
dimensionless coupling constant $\la$.
However, at the higher loop level, the decoupling of $\xi$ from $\b_\la$ is not obvious.
This is because $\xi$ is also the coupling constants of ${\phi'}^{\, n}$ in \eqref{action4}.
Therefore, in the following sections,
we discuss the decoupling of $\xi$ starting from the original form of the action \eqref{action}.

\section{Action \eqref{action} in terms of ordinary metric field}

We consider the ordinary perturbation theory of \eqref{action} with $\rho=0$
instead of introducing auxiliary field $\phi$.
We take the fluctuation field as
\bea
g_{\mu\nu}=\eta_{\mu\nu} +  h_{\mu\nu},
\eea
and  the gauge fixing term as
\bea
{\cal L}_{GF} =-\frac{1}{2a}\Big(\pa_\mu h^\mu{}_\nu-\frac{b}{2}\pa^\nu h^\a_\a \Big)^2,
\eea
where $a$ and $b$ are gauge fixing parameters.
We then find the quadratic part of the action is given by
\bea
{\cal L}^{(2)} &=& \frac14 h^{\mu\nu} \Big[ \left(\frac{1}{\la}\Box+\frac{M_P^2}{2} \right) P^{(2)}
+ \left(\frac{12}{\xi} \Box- M_P^2+ \frac{3 b^2}{2a} \right) P^{(0,s)} +\frac{1}{a} P^{(1)} \nn
&& \hs{20}
+ \frac{(b-2)^2}{2a} P^{(0,w)}+\frac{\sqrt{3}b(b-2)}{2a}(P^{(0,sw)}+P^{(0,ws)})
\Big]_{\mu\nu,\a\b} \Box\, h^{\a\b} .~~
\label{quad}
\eea
where
\bea
P^{(2)}_{\mu\nu,\a\b} &=& \frac{1}{2}\left(\t_{\mu\a}\t_{\nu\b}+\t_{\mu\b}\t_{\nu\a}
-\frac{2}{3}\t_{\mu\nu} \t_{\a\b}\right),\nn
P^{(1)}_{\mu\nu,\a\b} &=& \frac{1}{2}\left(\t_{\mu\a}\omega_{\nu\b}+\t_{\mu\b}\omega_{\nu\a}
+\t_{\nu\a}\omega_{\mu\b}+\t_{\nu\b}\omega_{\mu\a} \right),\nn
P^{(0,s)}_{\mu\nu,\a\b} &=& \frac{1}{3} \t_{\mu\nu}\t_{\a\b}, \qquad
P^{(0,w)}_{\mu\nu,\a\b} \,=\, \omega_{\mu\nu}\omega_{\a\b}, \nn
P^{(0,sw)}_{\mu\nu,\a\b} &=& \frac{1}{\sqrt{3}} \t_{\mu\nu}\omega_{\a\b}, \qquad
P^{(0,ws)}_{\mu\nu,\a\b} \,=\, \frac{1}{\sqrt{3}}\omega_{\mu\nu}\t_{\a\b},
\eea
are the spin projectors with
\bea
\t_{\mu\nu} = \eta_{\mu\nu}-\frac{\pa_\mu \pa_\nu}{\Box}, \qquad
\omega_{\mu\nu} = \frac{\pa_\mu \pa_\nu}{\Box}.
\eea
Then the propagator for the theory~\eqref{action} is given by
\bea
&& \frac{1}{(2\pi)^4} \Big[\frac{P^{(2)}}{k^2 \left[ \frac{1}{\la} k^2 - \frac{M_P^2}{2} \right]}
- \frac{a}{k^2} P^{(1)}
+\frac{P^{(0,s)}}{k^2 \left[ \frac{12}{\xi} k^2 + M_P^2 \right]}
-\frac{\frac{24 a}{\xi} k^2-3b^2+2a M_P^2}{(b-2)^2 k^2\left[\frac{12}{\xi}k^2+M_P^2\right]} P^{(0,w)}  \nn
&& \hs{20} -\frac{\sqrt{3} b}{(b-2)k^2\left[\frac{12}{\xi}k^2+M_P^2\right]} (P^{(0,sw)}+P^{(0,ws)})
\Big]_{\mu\nu,\a\b},
\label{prop}
\eea

This propagator becomes transparent in the Landau-type gauge, $a=b=0$, as
\bea
\frac{1}{(2\pi)^4} \Big[\frac{P^{(2)}}{k^2 \left[ \frac{1}{\la} k^2 - \frac{M_P^2}{2} \right]}
+\frac{P^{(0,s)}}{k^2 \left[ \frac{12}{\xi} k^2 + M_P^2 \right]}
\Big]_{\mu\nu,\a\b}.
\label{propL}
\eea
The first term represents the propagation of the traceless mode, and the second term
represents the propagation of the trace mode (or conformal mode).
At a first glance, both of the limits $\la\to\infty$ and $\xi\to\infty$ are dangerous because
the ultraviolet behavior is changed in both cases.
However, as we have discussed in section \ref{xitoinf}, the latter limit is safe.
In fact, by introducing an auxiliary scalar field $\phi$, the quartic behavior
of the second term in \eqref{propL} is resolved as
\bea
\frac{1}{k^2 \left[ \frac{12}{\xi} k^2 + M_P^2 \right]}=\frac{1}{M_P^2}\frac{1}{k^2}
-\frac{1}{M_P^2}\frac{1}{k^2 + \frac{\xi M_P^2}{12}}.
\eea
The system then becomes a coupled system of
the conformal mode and $\phi$, and the $\xi\to\infty$ limit is nothing but the large mass
limit for $\phi$. The important point is that this decoupling holds even when the interaction
is taken into account, as is discussed in section \ref{xitoinf}.

\section{ $\xi\to\infty$ limit and structure of bare action}

In this section we start with the action (\ref{action}), and examine the form of the counterterms
in the $\xi\to\infty$ limit.

As we have seen, the perturbation series is a double expansion with respect to $\la$ and $\xi$.
In particular the beta functions are degree $n+1$ polynomials of $\la$ and $\xi$ at the $n$-loop
level.
Because the simple pole ($1/\e$)  part of the counterterms is related to the beta functions
as (\ref{beta}) in the minimal subtraction scheme,
the $n$-loop counterterms must have the form
\begin{dmath}
S_{n\text{-loop}}=\sum_{k=1}^{n} \frac{1}{\e^k} \left[
\la^{n-1} \left(a^{(n) (k)}_1+a^{(n) (k)}_2\frac{\xi}{\la} + ... +a^{(n) (k)}_{n+2} \left(\frac{\xi}{\la}\right)^{n+1}\right)
 C_{\mu\nu\rho\la}^2
+\xi^{n-1} \left(b^{(n) (k)}_1+b^{(n) (k)}_2\frac{\la}{\xi} + ... +b^{(n) (k)}_{n+2} \left(\frac{\la}{\xi}\right)^{n+1}\right)
 R^2 \right].
\label{countern}
\end{dmath}
Each coefficient of $C_{\mu\nu\rho\la}^2$ and $R^2$ can contain
only up to $\la^{-2}$ or $\xi^{-2}$, respectively, in order
for the beta functions not to be singular in the zero coupling constant limit.
In particular, at the $1$-loop level we have
\bea
S_{1\text{-loop}}=\frac{1}{\e} \left[
\left(a_1+a_2\frac{\xi}{\la} + a_{3} \left(\frac{\xi}{\la}\right)^{2}\right) C_{\mu\nu\rho\la}^2
+\left(b_1+b_2\frac{\la}{\xi} +b_{3} \left(\frac{\la}{\xi}\right)^{2}\right) R^2
\right].
\label{counter1}
\eea

Now consider the limit $\xi \to \infty$.
If we assume that the bare action $S_0=S+S_{1\text{-loop}}+...$ remains finite in this limit,
we have
\bea
a^{(n) (k)}_2=...=a^{(n) (k)}_{n+2}=0, \qquad
b^{(n) (k)}_1=...=b^{(n) (k)}_{n-1}=0,
\eea
for all $n=1,...$ and $k=1,...,n$. In other words, in (\ref{countern}) only the first term
for $C_{\mu\nu\rho\la}^2$ and the last three terms for $R^2$ remain.
More explicitly, we can write as
\begin{dmath}
S_{n\text{-loop}}=\sum_{k=1}^{n} \frac{1}{\e^k} \left[
\la^{n-1} a^{(n) (k)} C_{\mu\nu\rho\la}^2
+\left(\la^{n-1}b^{(n) (k)}_n  +\xi^{-1}\la^{n} {b}^{(n) (k)}_{n+1}+\xi^{-2}\la^{n+1} {b}^{(n) (k)}_{n+2}\right)
 R^2 \right].
\label{counterreduced}
\end{dmath}
If this assumption is correct to all loop orders, from \eqref{beta} the beta functions should be given by
\bea
\beta_\la=\sum_{n=1}^{\infty} 4a^{(n) (1)} \la^{n+1}, \qquad
\beta_\xi=\sum_{n=1}^{\infty} 2 \left(b^{(n) (1)}_n \la^{n-1}\xi^2+{b}^{(n) (1)}_{n+1} \la^{n}\xi+{b}^{(n) (1)}_{n+2}
 \la^{n+1} \right).
\eea
Therefore, $\beta_\la$ is a function of only $\la$, and $\beta_\xi$ contains $\xi$ at most
quadratically. It will be interesting to see if this is correct at higher loop orders.

\section{Relation between various beta functions}

In the previous sections, we have seen that the action \eqref{action} is equivalent to \eqref{action4}
or \eqref{action5}.
Furthermore, if the cosmological constant is zero, $\rho=0$, $\xi$ is the mass squared for $\psi$.
Therefore, in the minimal subtraction scheme, the beta function for $\la$ does not depend on $\xi$,
which confirms the form of the first equation of \eqref{old}:
\bea
\b_\la^o = -\frac{1}{(4\pi)^2} \frac{133}{10}\la^2.
\label{betaS}
\eea

However, as we will discuss now, this beta function is not exactly the same as the beta function
for the decoupled action \eqref{Sinf} obtained by sending $\xi\to\infty$ in the action,
which we denote as $\beta_{\la}^{\infty}$.
In fact, it is known that $\beta_{\la}^{\infty}$ is given by~\cite{FT}
\bea
\b_\la^{\infty} = -\frac{1}{(4\pi)^2} \frac{797}{60}\la^2.
\label{betaSinf1}
\eea

The difference is exactly the contribution of a scalar field $\psi$.
Actually, for the action \eqref{action5} in the minimal subtraction scheme, the counterterm
does not depend on the mass of $\psi$. Therefore the beta function is the same as the massless
theory, that is, the theory with $\xi=0$. But this is nothing but the action \eqref{Sinf} plus
a massless scalar $\phi'$. The contribution of a scalar field  to the beta
function $\Delta\beta_\la$ is easily calculated from the induced Weyl gravity, and it is given by~\cite{FT,AM}
\bea
\Delta\b_\la = -\frac{1}{(4\pi)^2} \frac{1}{60}\la^2.
\label{betaSinf2}
\eea
And we can indeed find
\bea
\b_\la^o=\b_\la^{\infty}+\Delta\b_\la.
\label{betaSinf3}
\eea

It is now clear that the discrepancy is understood as the failure of the decoupling in the mass
independent renormalization scheme. In fact in the minimal subtraction scheme, the counterterm
remains the same even if the mass of a field becomes infinite and the field decouples physically.
Therefore the beta function for $\la$ of the original theory \eqref{action} in the minimal subtraction scheme
is not exactly the same as that of
the decoupled action \eqref{Sinf}, but we have an additional contribution from a scalar field.

Related to this is the difference of the coefficients of the beta functions directly
calculated in the conformal gravity~\cite{FT,BS1,OP2}, where the beta function is given as
\bea
\b_\la=-\frac{1}{(4\pi)^2}\frac{199}{15} \la^2,
\label{conf}
\eea
The difference between \eqref{betaSinf1} and \eqref{conf} is again the same as the contribution of
a scalar field~\eqref{betaSinf2}. This arises because \eqref{conf} is  calculated by projecting out
a trace mode in the theory. This suggests
that there might be some subtlety in the $\xi\to \infty$ limit related to the contribution of the spin 0 mode.
This was studied in \cite{OP2}, but the result remained the same. This point needs further study.

Finally what is the difference between the beta functions in the minimal subtraction scheme and
those in \eqref{new}?
The beta functions in \eqref{new} are read off from the scattering amplitudes.
In this case, there are additional contributions $\log(p^2/m^2)$ with an IR regulator $m^2$
when $M_P$ is set to zero. For the beta functions defined so, our simple observation may not apply because
such beta functions are not related to the counterterms.

%\newpage
\section{Summary}
\label{summary}

In this paper, we have studied the beta functions for the  dimensionless couplings  in the quadratic curvature
theory. By considering the general structure of the counterterms in the theory together with the $\xi\to\infty$
limit, we have argued that the beta function for $\la$ should be a monomial $\la^{n+1}$ and
that for $\xi$ consists of three terms of $\la^{n+1},\la^{n}\xi$ and $\la^{n-1}\xi^2$ at $n$-loop.
This result is supported by 1-loop calculation, but there has not been any calculation at 2-loop and beyond.
It would be an interesting problem to check explicitly whether this is indeed true or not  at 2-loop and beyond.
We also commented on the difference in the coefficients of several calculations of the beta functions for $\la$,
and there seems to be subtlety in the calculation. Sending $\xi\to\infty$ at the action level or
in the beta function gives difference in the mass-independent renormalization.
We also noted the different definition of the beta function given in \cite{BDMP}.

\section*{Acknowledgments}

We would like to thank Roberto Percacci for valuable comments.
H.K. thanks Prof. Shin-Nan Yang and his family for their kind support through the Chin-Yu
chair professorship. H.K. is partially supported by JSPS (Grants-in-Aid for Scientific Research
Grants No. 20K03970), by the Ministry of Science and Technology, R.O.C.
(MOST 111-2811-M-002-016), and by National Taiwan University.
The work of N.O. was supported in part by the Grant-in-Aid for Scientific Research Fund of the JSPS (C) No. 20K03980,
and by the Ministry of Science and Technology, R. O. C. (Taiwan) under the grant MOST 112-2811-M-008-016.
N.O. would like to thank Kimyeong Lee and Korea Institute for Advanced Study for their hospitality,
where this work was completed.

%\newpage

\end{document}